\documentclass{article}
\usepackage[utf8]{inputenc}
\usepackage{subcaption}
\usepackage{paralist}
\usepackage{hyperref}
\usepackage[export]{adjustbox}
\usepackage{listings}
\usepackage{tabularx}
\newcolumntype{Y}{>{\centering\arraybackslash}X}
\usepackage[bottom]{footmisc}
\usepackage{xcolor}

\usepackage[style=numeric]{biblatex}
\bibliography{references_nodoi} 

\newcounter{heuristic}
\newenvironment{heuristic}[1][]{\refstepcounter{heuristic}\par\medskip
   \noindent \textbf{Heuristic~\theheuristic~(#1).} \rmfamily\itshape}{\par\medskip}

\begin{document}

\title{{\bf Optimizing Federated Queries Based on the Physical Design of a Data Lake}\\{\color{red}work-in-progress paper}}

\author{Philipp D. Rohde, Maria-Esther Vidal \\
	TIB Leibniz Information Centre for Science and Technology  \\
	{\tt \{philipp.rohde,maria.vidal\}@tib.eu}
}

\date{}
\maketitle
\thispagestyle{empty}

\paragraph{\bf Abstract}
The optimization of query execution plans is known to be crucial for reducing the query execution time. 
In particular, query optimization has been studied thoroughly for relational databases over the past decades.
Recently, the \textit{Resource Description Framework} (RDF) became popular for publishing data on the Web.
As a consequence, federations composed of different data models like RDF and relational databases evolved.
One type of these federations are \textit{Semantic Data Lakes} where every data source is kept in its original data model and semantically annotated with ontologies or controlled vocabularies.
However, state-of-the-art query engines for federated query processing over Semantic Data Lakes often rely on optimization techniques tailored for RDF.
In this paper, we present query optimization techniques guided by heuristics that take the physical design of a Data Lake into account. The heuristics are implemented on top of \textit{Ontario}, a SPARQL query engine for Semantic Data Lakes.
Using source-specific heuristics, the query engine is able to generate more efficient query execution plans by exploiting the knowledge about indexes and normalization in relational databases.
We show that heuristics which take the physical design of the Data Lake into account are able to speed up query processing.

\paragraph{\bf Keywords}
federated query processing, data lake, database physical design, query optimization

\section{Introduction}
Advances in the technologies for data generation and ingestion facilitate the collection of large volumes of data from where valuable knowledge can be extracted.
However, the wide variety of formats and data management systems available for storing and processing the collected data, hamper interoperability and data integration.
The problem of integrating data collected from different data sources has been extensively treated in the literature~\cite{0029346,HalevyRO06};
the mediator and wrapper architecture proposed by Wiederhold~\cite{Wiederhold1992} and the data integration system approach presented by Lenzerini~\cite{Lenzerini02}, represent the basis for the state of the art in data integration~\cite{GolshanHMT17,KnoblockS15,Mountantonakis:2019:LSI:3362097.3345551} and query processing over heterogeneous data sets or polystores~\cite{Duggan:2015:BPS:2814710.2814713,Endris2019,HaiGQ16,Khan2019,QuixHV16}.
Albeit the rich variety of solutions, the problem of efficiently querying heterogeneous data srouces remains still open because data sources may differ in many various parameters, e.g., the physcial implementation of the databases that store the data.
In order to effectively solve interoperability and take advantage of the huge amount of available data, novel query processing solutions able to exploit not only logical characteristics of the data but also their physical representation, are demanded. 

\begin{figure*}[t]
    \centering
    \begin{subfigure}[b]{0.41\textwidth}
        \centering
        \raisebox{25mm}{\includegraphics[width=\linewidth,valign=c]{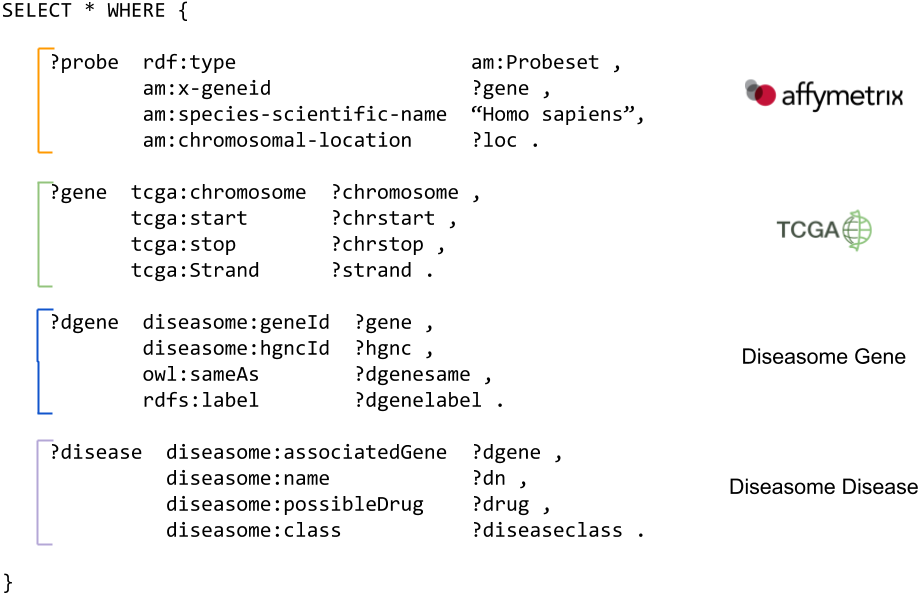}}
        \subcaption{SPARQL Query}
        \label{fig:mot:query}
    \end{subfigure}
    \hfill
    \begin{subfigure}[b]{0.28\textwidth}
        \centering
        \raisebox{25mm}{\includegraphics[width=\linewidth,valign=c]{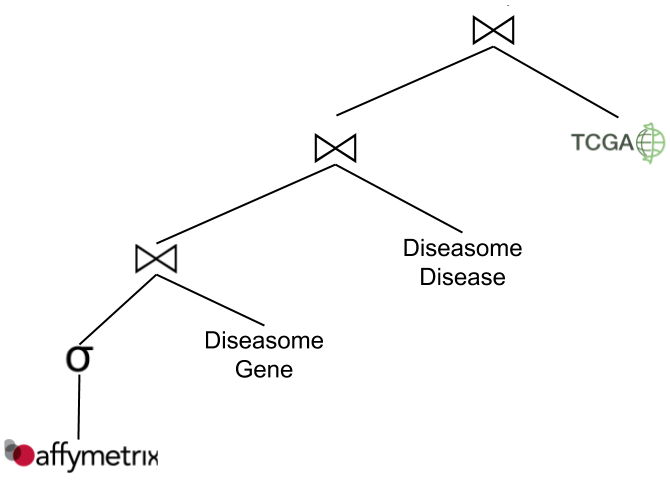}}
        \subcaption{QEP without using indexes}
        \label{fig:mot:woidx}
    \end{subfigure}
    \hfill
    \begin{subfigure}[b]{0.28\textwidth}
        \centering
        \raisebox{25mm}{\includegraphics[width=\linewidth,valign=c]{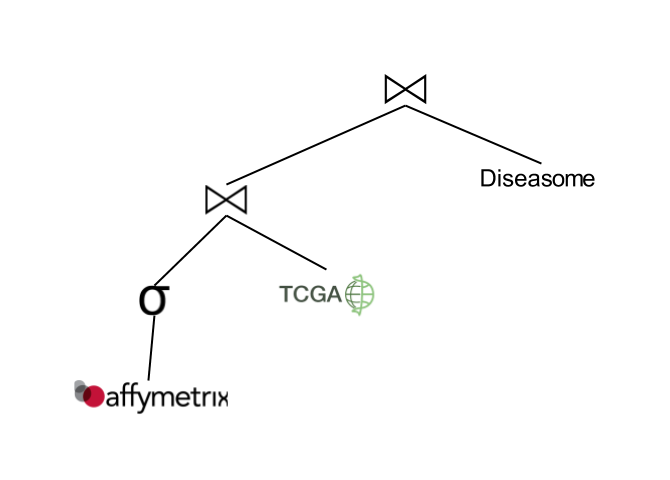}}
        \subcaption{QEP using indexes}
        \label{fig:mot:idx}
    \end{subfigure}
  \caption{{\bf Motivating example.} Query execution plans (QEP) for the same query (a); not considering indexes (b) and considering indexes (c). Optimizing QEPs with respect to the physical design of the Data Lake allows to find more efficient plans with fewer operations needed to be performed at the query engine level.}
  \label{fig:mot}
\end{figure*}

\paragraph{\bf Motivation.}
Considering the query in \autoref{fig:mot:query} two different query execution plans (QEP) can be generated.
On the one hand, the QEP in \autoref{fig:mot:woidx} is unaware of the physical design.
Therefore, as many operations as possible are performed at the level of the query engine.
On the other hand, the QEP in \autoref{fig:mot:idx} is aware of the physical design.
Hence, as many operations as possible are pushed to the data sources.
In the example query, the information about genes and diseases is from the \textit{Diseasome} data set and stored in a single source.
Therefore, the join can be pushed down.
The filter expression for the scientific name of the species in the \textit{Affymetrix} data set is always performed at the query engine because it is not indexed.
No index is created since there are values that are present in more than 15\% of the records.


This paper is organized as follows.
Preliminary concepts and heuristics for optimizing federated queries are discussed in Section 2.
Section 3 provides a preliminary analysis.
Related work is presented in Section 4.
We conclude in Section 5.

\section{Our Approach}
\subsection{Preliminaries}
In this section, we present basic concepts required to understand this work.
SPARQL queries can be partitioned into groups of acyclic patterns that share exactly one variable~\cite{Vidal2010}.
In the common case they represent a class of instances that share the same properties.
Decomposing a query based on these groups leads to a QEP with \textit{star-shaped sub-queries} (SSQ) in the leaves.
In the motivating example (cf. \autoref{fig:mot:query}) the SSQs are indicated with colored brackets.
Following the idea of star-shaped groups over subjects \textit{RDF Molecule Templates} (RDF-MT)~\cite{Endris2018} are an abstract description of the properties of the entities in an RDF data set.
Each RDF-MT represents one class of instances, e.g., drugs from Diseasome or genes from TCGA.
A \textit{Data Lake} is a collection of heterogeneous data sets.
The data sets do not necessarily share the same data model.
If data models that do not have semantics by nature, e.g., relational databases, are annotated with semantics, the collection of data sets is called \textit{Semantic Data Lake}.
RDF and relational databases are amongst the most frequent data models present in Semantic Data Lakes.
In our case, the query engine receives a SPARQL query and translates sub-queries to the native query language, e.g., SQL, of the data source.

\subsection{Source-Specific Heuristics}
One problem to tackle during query processing over a Data Lake is the variety of data models used throughout the Data Lake.
State-of-the-art query engines use generalized optimization techniques or rely on heuristics tailored for one specific data model.
Hence, they lose further opportunities for improving the query performance.
In order to enable the maximal capability of optimizing the query execution plans, the physical design of the Data Lake needs to be considered.
This includes optimizing sub-queries for the different data models present in the Data Lake.
We propose two heuristics designed for relational databases to show the impact of respecting the physical design.
The proposed heuristics assume that the relational tables are normalized in 3NF.
Further we expect that the subjects of a SPARQL query are modeled as the primary keys of the tables.
Jozashoori and Vidal~\cite{Jozashoori2019} showed that this is the best case scenario for running star-shaped sub-queries against relational databases.

\begin{heuristic}[Pushing down joins]\label{heu:join}
    Given two star-shaped sub-queries over the same RDB endpoint, combine those sub-queries into one sub-query if the join attribute is indexed.
\end{heuristic}

\autoref{heu:join} is proposed since joins over indexed attributes in relational databases are considered to be fast as long as the number of joins is kept reasonable.
In the data we are working on, star-shaped sub-queries are usually represented by less then four relational tables.
Assuming the worst case for a star-shaped sub-query, three relational tables contribute to the answer.
Two of those tables are connected with the remaining table via foreign keys.
Therefore, joining two star-shaped sub-queries leads to a six-way join in the worst case.
Hence, the number of joins can be considered as reasonable.
In order to decide if the number of joins is kept reasonable, a later version should consider the number of relational tables involved.
Not only the join performance of RDB engines justifies this heuristic but also the possibility of a reduced size of the intermediate result.
\autoref{heu:join} improves the query performance by reducing the time needed to perform the join as well as possibly decreasing the intermediate result. 

\begin{heuristic}[Pushing  up instantiations]\label{heu:select}
    Given a star-shaped sub-query over a relational database, perform filters on query engine level unless there is an index on the filtered attribute and the network speed is low. 
\end{heuristic}

From our experience filtering string data at the query engine performs faster compared to executing the filters in the relational database.
Therefore, \autoref{heu:select} is expected to speed up the query execution, 
even though a larger intermediate result has to be transferred to the query engine. 
However, if the network speed is low, the intermediate result has to be minimized and, therefore, the instantiation is performed at the relational database.
\autoref{heu:select} leads to faster query execution through speeding up the filter evaluation in case of fast networks at the cost of transferring larger intermediate results.

The proposed heuristics follow common knowledge about relational databases and network delays.
Relational databases are designed to find effective and efficient query execution plans for joins and filter expressions exploiting indexes if beneficial.
Even if the execution time at one source is increased by combining sub-queries into one, the overall query performance might be improved in the case of a slow network by reducing the intermediate result.
Hence, the heuristics are very well suited to investigate the impact of considering the physical design of the Data Lake during the query optimization.

\section{Experiment}
We empirically study two different kinds of query execution plans in order to evaluate the proposed heuristics.
The QEPs are as follows:
\begin{inparaenum}[\bf a\upshape)]
    \item \emph{Physical-Design-Unaware QEP}: A QEP not respecting the physical design of the Data Lake and, therefore, not using the generated indexes to optimize the query execution.
    \item \emph{Physical-Design-Aware QEP}: A QEP that considers the indexes present in the relational database.
\end{inparaenum}
The source code and the data used for the experiment as well as the results are available at GitHub\footnote{\url{https://github.com/SDM-TIB/Ontario-SEAData2020}}.

\begin{figure*}[t]
    \centering
    \begin{subfigure}[b]{0.32\textwidth}
        \centering
        \includegraphics[width=\linewidth]{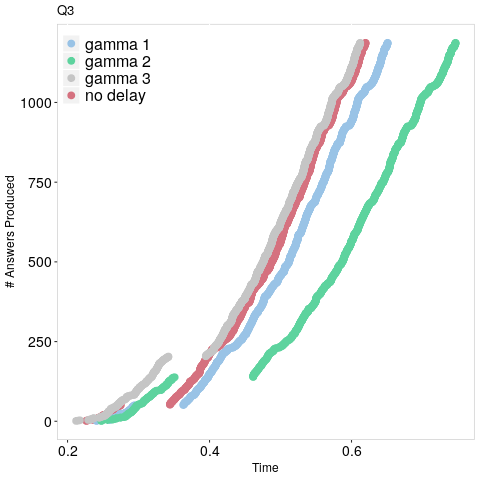}
        \subcaption{Physical-Design-Unaware QEPs}
        \label{fig:traces:woidx}
    \end{subfigure}
    \hfill
    \begin{subfigure}[b]{0.32\textwidth}
        \centering
        \includegraphics[width=\linewidth]{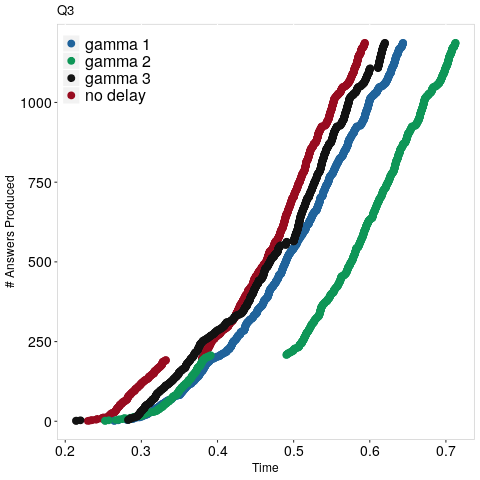}
        \subcaption{Physical-Design-Aware QEPs}
        \label{fig:traces:idx}
    \end{subfigure}
    \hfill
    \begin{subfigure}[b]{0.32\textwidth}
        \centering
        \includegraphics[width=\linewidth]{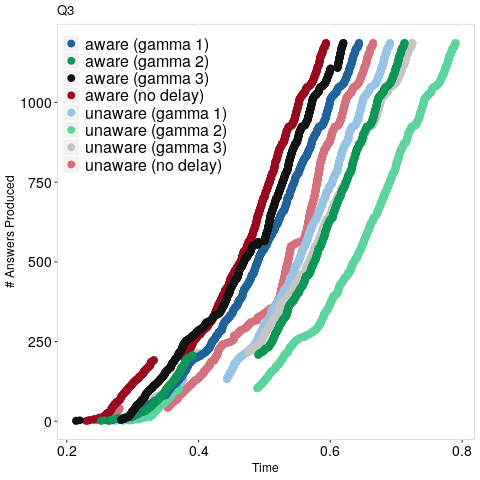}
        \subcaption{both QEPs}
        \label{fig:traces:both}
    \end{subfigure}
  \caption{{\bf Answer Traces for Q3} with no delay and three different delays according to the gamma distributions. Answer traces show the generation of answers over time (in seconds). (a) Physical-Design-Unaware QEP not using indexes; (b) Physical-Design-Aware QEP using indexes whenever possible; and (c) both QEPs in comparison. Slow networks have a higher impact on physical-design-unaware QEPs. Respecting the physical design improves query performance.}
  \label{fig:traces}
\end{figure*}



\paragraph{\bf Data Sets.}
In this experiment, we use the data sets from the LSLOD benchmark~\cite{Hasnain2017} which is composed of ten real-world data sets from the life sciences domain of the Linked Open Data (LOD) cloud.
The RDF version of each data set is transformed into relational tables.
These tables are then normalized to 3NF.
Indexes are created for the primary keys.
Furthermore, additional indexes for some attributes that are used for joins or selections in the queries used are generated to evaluate the impact of the proposed heuristics.
The data of each LSLOD data set are uploaded into a dedicated \emph{MySQL 5.7.24} Docker container.

\paragraph{\bf Queries.}
The queries provided for the LSLOD benchmark do not contain the possibility of pushing down the join of two star-shaped sub-queries.
Therefore, we do not use the provided queries and created five queries tailored for the heuristics to show their impact on query performance.
The following parameters were considered during the development of the new queries:
\begin{inparaenum}[\bf a\upshape)]
    \item the selectivity of the query,
    \item filter expressions over indexed attributes, and
    \item possible joins of star-shaped sub-queries over indexed attributes.
\end{inparaenum}
Another factor that impacts on the performance of a query is the size of the intermediate result.

\paragraph{\bf Network Simulation.}
We used four different network settings which simulate the following networks:
\begin{inparaenum}[\bf a\upshape)]
    \item \textit{No Delay}: perfect network with no or negligible latency.
    \item \textit{Gamma 1}: fast network with a gamma distribution ($\alpha = 1, \beta = 0.3$) of response latency resulting in an average latency of 0.3 milliseconds.
    \item \textit{Gamma 2}: medium fast network with an average latency of 3 millisecons resulting from a gamma distribution ($\alpha = 3, \beta = 1$).
    \item \textit{Gamma 3}: slow network with a gamma distribution ($\alpha = 3, \beta = 1.5$) leading to an average latency of 4.5 milliseconds per message.
\end{inparaenum}

\paragraph{\bf Setup.}
For the purpose of the experiment, Ontario~\cite{Endris2019} was modified to run physical-design-aware QEPs and physical-design-unaware QEPs.
Network delays are simulated within the SQL wrapper of Ontario; delaying the retrieval of the next answer from the source.
The duration of the delay is calculated using the {\tt numpy.random.gamma()} function and the delay is produced using the Python {\tt time.sleep()} function.
Like the data sources, Ontario is running in a Docker container.
All Docker containers were running at the same server.
Hence, network costs other than introduced by the network simulation can be neglected. 
The experiments were executed on an Ubuntu 16.04.6 LTS 64 bit machine with two Intel(R) Xeon(R) Platinum 8160 2.10 GHz CPUs
, and 755 GiB DDR4 RAM.


\paragraph{\bf Preliminary Results.}
The experiment conducts of eight different configurations in total, i.e., both QEP types are evaluated using all four simulated network conditions.
In doing so, we enable analyzing the impact of different network conditions and not only the impact of physical-design-aware execution plans.
The analysis shows that the impact of network delays is higher in the case of physical-design-unaware query execution plans.
An analysis of the results suggests that the proposed heuristics have potential to improving the query performance.
However, the heuristics need to be evaluated more thoroughly and revised.
The evaluation of \autoref{heu:join} is currently limited due to the query translation of Ontario.
The translation of SPARQL queries into SQL queries is not optimized for combining star-shaped sub-queries.
This leads to an increase in the query execution time if the join is pushed down.
Forcing Ontario to send the optimized SQL query for Q2 approx. halves the execution time compared to the physical-design-unaware QEP.

Even though \autoref{heu:select} seems to be correct from our experience, a deeper study on the difference of the filter execution performance between relational database and query engine is needed.
On the one hand, the results of Q1 support our experience and suggest to follow \autoref{heu:select}.
On the other hand, the results of Q3 suggest otherwise.
\autoref{fig:traces} shows the answer generation for Q3 over time.
It can be seen that executing the filter at the relational database (physical-design-aware QEP) is faster for this query.
Therefore, more studies on the filter execution need to be done.
Additionally, the experiment shows that the proposed heuristics are impacted by the implementation of other optimizations that are performed by Ontario.
The impact of the heuristics on the query performance is not only influenced by the physical design of the Data Lake and the network conditions but also by the implementation of the query engine and wrappers.

\section{Related Work}
\paragraph{\bf Federated Databases.}
The problem of integrating data from dissimilar data sources has been addressed in the literature by implementing the mediator and wrapper architecture proposed by Wiederhold~\cite{Wiederhold1992}.
Several federated query engines have been defined in the context of relational database~\cite{Florescu1998,Halevy2001,Ives2004,Zadorozhny2002}, as well as diverse of integration frameworks~\cite{Halevy2006}.
We focus mainly on approaches that implement strategies to address the problem of source selection and decomposition of SPARQL queries, although, we recognize the tremendous advance that the Database community has done to the general problem of data integration in the last fifteen years.
Existing approaches are grouped according to the amount of knowledge that describes the data sources, and that is exploited during source selection and query decomposition to enhance the quality of the generated query decompositions.

\paragraph{\bf Federations of RDF}
With the rise of the Resource Description Framework (RDF) new federated query engines were proposed to optimize query processing over the new data model.
\emph{FedX}~\cite{Schwarte2011} is one of those query engines.
FedX aims at minimizing the number of requests to be sent to the sources by identifying groups of triple patterns that can be exclusively evaluated by a single endpoint.
\emph{ANAPSID}~\cite{Acosta2011} stores a list of predicates that each endpoint is able to answer.
Queries are decomposed into star-shaped sub-queries.
ANAPSID introduces adaptive physical operators to generate results as soon as they arrive from the sources.
These operators perform better than the traditional blocking operators.
MULDER~\cite{Endris2018} is based on ANAPSID and describes the sources in terms of RDF Molecule Templates (RDF-MTs).
MULDER is able to reduce the query execution time and increase answer completeness by using semantics in the source descriptions during decomposition and source selection.
Query engines for federations of RDF sources can benefit from the semantics of the metadata and received data.

\paragraph{\bf Polystores.}
More recently, the research focus shifted towards query processing against heterogeneous data sources.
Different approaches have been proposed on how to store, integrate, and query heterogeneous federations.
\emph{SeBiDA}~\cite{Mami2016} is a proof-of-concept for a semantified big data architecture.
Data sets are differentiated in semantic, annotated with semantics, and non-semantic.
The latter can optionally be lifted with semantics if mappings are provided.
SeBiDA uses \emph{Apache Spark} to reformat the data according to classes.
The data is reformatted in \emph{Apache Parquet} tables.
Therefore, the data is integrated in a centralized or clustered manner and can be queried using SQL.
Contrary to SeBiDA, \emph{PolyWeb}~\cite{Khan2019} and \emph{Ontario}~\cite{Endris2019} keep the data sources in their original data model.
Data sources are queried in their native query language while the user sends SPARQL queries to the query engine.
PolyWeb uses the same cost-based model as FedX does and \textit{predicate-based join groups} to reduce the number of local joins.
Other than PolyWeb, Ontario is based on MULDER and uses the same plan generator extended with heuristics for better optimization potential.
Ontario also uses the physical operators of ANAPSID.
Several query processing engines have been proposed, but most of them focus on a single data model for query optimization and therefore miss optimization opportunities.

\section{Conclusion}
In this paper we present two rather simple heuristics that aim at improving the query performance by considering the physical design of the Data Lake compared to state-of-the-art query execution plans.
Our heuristics take the
\begin{inparaenum}[\bf a\upshape)]
    \item presence of indexes, and
    \item network condition
\end{inparaenum}
into account.
Even though the heuristics and their implementation are in an early stage, we can conclude that the query performance in a Data Lake can be improved when considering the characteristics of each data model.
In future work, we plan to overcome the described limitations, e.g., improving the quality of the translated SQL queries.
Furthermore, we will investigate the performance of different implementations of relational databases in order to gain a deeper understanding of why filter expressions seem to perform better at query engine level in most cases even though the intermediate results are larger in that case.
Additionally, studying different kinds of query decomposition (e.g., triple-based instead of star-shaped sub-queries) and not normalized tables is part of our plans.

\paragraph{\bf Acknowledgments}
This work has been partially supported by the EU H2020 RIA funded projects QualiChain (No 822404) and iASiS (No 727658).

\printbibliography

\end{document}